\documentclass[%
 reprint,
 amsmath,amssymb,
 aps, prd,
 twocolumn,
]{revtex4-2}

\usepackage{graphicx}
\usepackage{dcolumn}
\usepackage{bm}
\usepackage{xcolor}
\usepackage{url}
\usepackage{ulem}
\usepackage{hyperref}
\hypersetup{
     colorlinks = true,
     linkcolor = blue,
     anchorcolor = blue,
     citecolor = green,
     filecolor = blue,
     urlcolor = magenta
     }

\begin{document}


\title{Effective chirp mass in the inspiral frequency evolution of the non-spinning eccentric compact binary}

\author{Nirban Bose}
\email{nirban@iitb.ac.in}
\author{Archana Pai}%
 \email{archanap@iitb.ac.in}
\affiliation{%
Department of Physics, Indian Institute of Technology Bombay, Mumbai, Maharashtra 400076, India
}%

\date{\today}

\begin{abstract}
Compact binary systems with black holes are the primary sources of interferometric advanced gravitational wave detectors.  
Astrophysical models suggest possibility of binary systems with appreciable non-zero eccentricity in the dense stellar environment like globular clusters and galactic nuclei. While most of the compact binary events have not shown appreciable eccentricity, constrains on the eccentricity have been placed on few detected events. With increasing sensitivity of the gravitational wave detectors, the eccentric binaries are plausible sources in the gravitational wave window.
Nevertheless, the challenges in the waveform modelling for high values of eccentricity constrain both the search methodologies of eccentric system as well as parameter estimation of the eccentricity. Waveform independent approaches are being investigated. In this work, we demonstrate that a new {\it effective chirp mass $\mathcal{M}_e$} parameter can be defined which governs the inspiral dynamics of the non-spinning eccentric compact binary system. We obtain the phenomenological model of the same for low to moderate eccentricity values. One direct implication is its application in the unmodelled searches and as an another implication, we demonstrate that this model can be used to constrain the eccentricity of the non-spinning eccentric binaries in the advanced detector era in the waveform model-independent way. 
\end{abstract}


\maketitle






\section{Introduction}

Gravitational wave(GW) astronomy has grown rapidly with the detection of gravitational wave signal from the merger of compact binary objects like binary black holes, binary neutron stars and neutron star black hole systems by the advanced ground based GW detectors like Advanced LIGO \textcolor{green}{\cite{TheLIGOScientific:2014jea}}, Virgo~\cite{TheVirgo:2014hva}. So far, the observational runs have yielded $\sim$ 50 confirmed detections of compact binary coalescence mergers \cite{PhysRevX.9.031040, PhysRevX.11.021053, PhysRevLett.125.101102,GW150914-DETECTION,GW170608-DETECTION,GW170817-DETECTION,LIGOScientific:2020stg,Abbott:2020khf}. With continuous improvement in the sensitivity of the detectors and additional detectors like KAGRA \cite{Akutsu:2020his} and LIGO India \cite{Ligo_India,saleem2021science} coming up, we expect to observe many more compact binary merger events.  While most of the observed detections do not show signature of eccentricities, a number of studies suggest that compact binaries formed in the dense stellar environment 
can have a significant fraction of them with non-negligible eccentricities \textcolor{green}{\cite{Rodriguez:2017pec}}. In fact, there has been recent studies to constrain the eccentricities of some of the detected events \cite{Gayathri:2020coq, Romero_Shaw_2020,Nitz:2019spj,Romero_Shaw_2019}. 
With increasing sensitivity of the ground based GW detectors, they show detection prospects in the sensitive band of the detectors. Thus, eccentric binaries are plausible sources of GW in the upcoming observational runs that can carry their formation channels' signature. 

Unlike isolated binary systems,  eccentric binaries are generally nurtured in dense stellar environments like globular clusters, galactic nuclei etc. \textcolor{green}{\cite{2019ApJ...871...91Z}}. In these dense stellar clusters, dynamical encounters and close fly bys between binary systems and binary-single systems are quite common. If the encounter is strong enough during such encounters, it may knock the binary out of the cluster and lead to merger. If the knock is not strong enough, the binary will get \textit{hardened} by accumulating more such \textit{soft} encounters and ultimately merge by emitting GW \cite{Samsing_2014}. In either case, the binaries do not get enough time to get circularized by the time they enter the detector band and retain a non negligible eccentricity. Often there can also be scenarios where a third body can perturb the orbit of the inner binary and induce periodic oscillations of the eccentricity of the binary, a phenomena known as Kozai-Lidov oscillations \textcolor{green}{\cite{Wen_2003}}. In the case of galactic nuclei, a central supermassive black hole plays the role of the perturber and induces the Kozai-Lidov oscillations. Different eccentricity distributions characterize all these different formation channels of the binaries. Probing the eccentricity of these binaries would provide information about the different formation channels of the binary black holes. 

The existing model-based matched filter searches do not have a suitable template bank to detect eccentric binary black hole (eBBH) systems. There have been efforts to develop template based searches for eccentric binary neutron star systems for dominant mode GW frequency but limited by eccentricity range \cite{Nitz_2020}. The lack of template based searches for eBBH systems can be attributed to the lack of accurate waveform models encompassing a wide range of eccentricity, spins, mass ratios, etc. Several eccentric waveforms include EccentricTD \textcolor{green}{\cite{Tanay:2016zog}}; a time-domain, inspiral only, non-spinning waveform model valid for high eccentricities and up to second order post-Newtonian corrections. EccentricFD (\textcolor{green}{\cite{Huerta:2014eca}}), on the other hand is an analytic, frequency-domain, inspiral only, non-spinning model valid for eccentricities up to $0.4$. It reduces to the quasi-circular post-Newtonian approximant TaylorF2 at zero eccentricity limit. The SEOBNRE \textcolor{green}{\cite{PhysRevD.96.044028}} waveform is based on the effective-one-body formalism and numerical relativity simulations. It is a full inspiral-merger-ringdown waveform valid for eccentricities up to $0.2$ and aligned spin systems. Another time-domain, the non-spinning, full inspiral-merger-ringdown model is the ENIGMA model \textcolor{green}{\cite{PhysRevD.97.024031}}, valid for low eccentricities $\leq 0.2$. It is developed using post-Newtonian theory, self-force, and black hole perturbation theory. Apart from these, there are also several numerical relativity waveforms valid for discrete values of eccentricities, spins, mass ratios, etc \textcolor{green}{\cite{Boyle_2019}}. As we note, all of them have their limitations in terms of eccentricity range as well as spins. This further limits the detailed parameter estimation study involving the complete range of the parameters. However, there are ongoing efforts to develop complete inspiral-merger-ringdown waveforms, including spins, precession up to high eccentricity.  Current eBBH searches rely on the model independent approach, based on excess power methods, which was used in search of eccentric systems in the first two observational runs \cite{Abbott_2019}.
Thus, it is crucial to develop robust, model independent methodologies for detection and constrain the parameters of the eccentric compact binary system.

In this work we take one step forward in that direction. We know that the chirp mass $\mathcal{M}$; a combination of reduced mass and the total mass captures the GW frequency evolution (quadrupolar (2,2) mode) from the quasi-circular binary system. Here, we obtain an equivalent new {\it effective chirp mass} parameter in case of eccentric  systems;
an {\it effective chirp mass} parameter which captures the frequency evolution of (2,2) mode, excluding the eccentric higher harmonics, of the non-spinning eccentric binary. We provide a phenomenological model for the effective chirp mass parameter in terms of the chirp mass and the eccentricity at the reference frequency. We further discuss the implication of this model in the model-independent framework.

The paper is organized as follows, in section \ref{orbital dynamics}, we give a background of the orbital dynamics of non-spinning eccentric systems and in \ref{ft-evolution} we discuss  frequency evolution and {\it effective chirp mass} of the systems. In section \ref{model}, we obtain the phenomenological model for the {\it effective chirp mass} of non-spinning eccentric system. In section \ref{Eccentric non-spinning inspiraling binary in the advanced detector}, we study SNR of the astrophysically motivated eccentric systems 
in the advanced LIGO-like detector. In section \ref{MeFromQtransform}, We study the implication of the {\it effective chirp mass} while obtaining the chirp mass from the time frequency based f-t evolution in the advanced detector era. In Sec. \ref{Implication on the chirp mass and eccentricity estimation} we study the implication on obtaining constrains on the eccentricity and chirp mass parameter.
Finally, in section \ref{conclusions}, we summarize our work and discuss the future plans.

\section{GW frequency evolution of non-spinning eccentric system}
\label{Gravitational wave frequency evolution of non-spinning eccentric system}

According to GR, any compact binary system in an eccentric orbit emits GW following the quadrupole moment formula. While the seminal work was done in \cite{peters}, a volume of work focused on modelling the gravitational waveform from an eccentric system exists \cite{Yunes_2009,Arun:2007rg,PhysRevD.99.124008}. Here, our focus in not the waveform modelling rather, we focus on the GW frequency evolution of the quadrupolar non-eccentric part of the (2,2) mode for the non-spinning eccentric compact binaries in the inspiral phase. The primary motivation is to study the effect of eccentricity on this frequency evolution.

\subsection{Keplerian dynamics}
\label{orbital dynamics}

Under the quadrupole and adiabatic approximation, the semi major axis ($a$), the eccentricity of the orbit ($e$), the gravitational energy ($E$), and the frequency of emission of GW ($f$) varies slowly over the orbital period. Using the Keplerian orbits, the average rate of change of the quantities are analytically calculated in \cite{peters,PhysRev.136.B1224}.
Compared to circular systems, eccentric binaries have higher asymmetry associated with them and thus the rate of emission of GW energy is higher for the eccentric system, so does the rate of instantaneous GW frequency as shown below:
\begin{subequations}
\begin{align}
&\bigg<\frac{da}{dt}\bigg> = -\frac{64 G^3 }{5 c^5 } \frac{\mu {M}^2}{a^3 {(1-e^2)}^{7/2}} \left[1+ \frac{73e^2}{24}+\frac{37e^4}{96} \right] \label{da_dt}\\
   &\bigg<\frac{dE}{dt}\bigg> = -\frac{32 G^4 \mu^{2} {M}^3}{5 c^5 a^5 (1-e^2)^{7/2}} \left[1+ \frac{73e^2}{24} + \frac{37e^4}{96} \right] \label{dE_dt}\\
   &\bigg<\frac{de}{dt}\bigg> =-\frac{304 G^3}{15 c^5} \frac{ \mu {M}^2 e }{a^4 {(1-e^2)}^{\frac{5}{2}}} \left[1+\frac{121 e^2}{304} \right] \label{de_dt}\\
   &\bigg< \frac{df}{dt} \bigg> = \frac{96 {\pi}^{8/3}}{5 c^5}  \frac{ {(G\mathcal{M})}^{5/3} f^{11/3}}{{(1-e^2)}^{7/2} } \bigg[1 + \frac{73 e^2}{24} + \frac{37 e^4}{96}  \bigg] \label{df_dt} 
   \end{align}
   \label{Peters and Mathews equations}
 \end{subequations}\\
 where $\mathcal{M} = \mu^{3/5} M^{2/5}$ is the chirp mass of the system in terms of the reduced mass $\mu$ and total mass $M$ of the system. The $G,c$ are the universal gravitational constant and speed of light respectively. The angular brackets represent the time averaged quantities over one cycle. These equations are non-linear coupled equations and their post-Newtonian (PN) corrected version has been used to obtain the time-domain waveforms such as EccentricTD. 
In EccentricTD\textcolor{green}{\cite{Tanay:2016zog}}, the authors compute the time domain waveform using an accurate and efficient prescription which incorporates the orbital eccentricity into the quasi-circular time-domain TaylorT4 approximant at 2PN order. This approach is an extension of \textcolor{green}{\cite{PhysRevD.70.064028}} where $h_+$ and $h_{\times}$ for compact binaries in eccentric orbits as a sum over eccentric harmonics.

 \subsection{$f-t$ evolution and effective chirp mass $\mathcal{M}_e$}
\label{ft-evolution}
In this subsection, we obtain the f(t) evolution of the eccentric system without the higher eccentric harmonics in the quadrupolar (2,2) mode at the Newtonian order using the Keplerian dynamics. Henceforth we refer to this derived frequency evolution as f(t) even though it does not include the higher eccentric harmonics in the evolution. We further this point in the conclusion. As a first step, we use Eqn (\ref{da_dt}) and (\ref{de_dt}), and integrate $da/de$ under the initial conditions as at
$a=a_0, e=e_0$ \cite{peters}. We use the Kepler's law to obtain in integration constant assuming that at the initial condition $f=f_L$ and is given as (see Appendix \ref{appendix B}),
\begin{equation}
    a(e) = \frac{{(G {M})}^{1/3} {(\pi f_L)}^{-2/3}  e^{12/19}}{g (e_0) (1-e^2)}\bigg[1+{\frac{121 e^2}{304}}\bigg]^{870/2299} \,.
    \label{a(e)}
 \end{equation}
Here, the $f_L$ is the lower cut off frequency of the detector and the $g(e)$ function is 
\begin{equation}
\label{ge}
g(e)=\frac{e^{12/19}}{(1-e^2)}\bigg[1+{\frac{121 e^2}{304}}\bigg]^{870/2299} \,.
\end{equation}
 
We substitute $a(e)$ in  the R.H.S. of Eq. (\ref{de_dt}) and obtain 
\begin{equation}
\bigg<\frac{de}{dt}\bigg> =-\frac{304 g^4(e_0)}{15 c^5}\frac{{(G\mathcal{M})^{5/3}}  e^{-29/19}}{{(\pi f_L)}^{8/3}{(1-e^2)}^{-3/2}} {\left[1+\frac{121 e^2}{304} \right]}^{\frac{-1181}{2299}}
\label{de_dt2}
\end{equation}
Under the slow motion, adiabatic approximation, the rate of change of eccentricity is a function of chirp mass and eccentricity. This is also consistent to the fact that the mass-ratio enters in the 1 PN order correction in the phase and not in the Newtonian order. 
 
We numerically evolve the eccentricity $e$ as a function of time using the RK4 method with the chosen initial conditions. The signal enters the detector band at $t=t_0$ when the GW frequency coincides with $f_L$.
The corresponding eccentricity is denoted as $e_{f_L}$. 
Next, we use the eccentricity obtained at each time instant with the fixed time interval in the R.H.S. of Eqn (\ref{df_dt}) and obtain the frequency at the next time instant.
Thus evolving at each time step, we get, $f(t)$ and $e(t)$.
  
\begin{figure}
\includegraphics[scale=0.5]{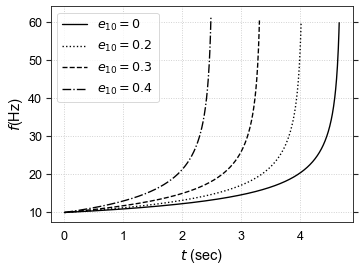} 
\caption{$f(t)$ evolution for equal mass systems with total mass of 70 $M_{\odot}$ and eccentricities 0 (solid), 0.2 (dotted), 0.3 (dashed) and 0.4 (dash-dotted). The inspiral signal duration is 4.7, 4.0,3.0 and 2.5 seconds respectively. Frequency evolves as a power law in case of a circular case.}
\label{ftplot}
\end{figure}

We consider GW150914 like binary system with equal component masses as $35 M_{\odot}$ with $f_L=10 Hz$ and corresponding eccentricities $e_{10}=0,0.2,0.3,0.4$. We evolve the system in the inspiral phase and obtain $f(t)$ as described above. Fig.\ref{ftplot} shows this frequency evolution up to $f_{LSO}$. The binary systems with higher eccentricity merge quickly owing to its higher power emitted, and thus shorter is the time spent in the detector band. 
 
\begin{figure}
\begin{tabular}{c}
\includegraphics[scale=0.5]{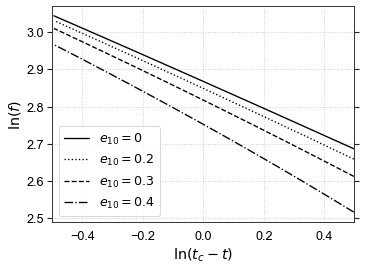} \label{logf-logt diff e}
\end{tabular}
\caption{The $ln(f)-ln(t_c-t)$ variation primarily in the inspiral part for equal mass systems with total mass of 70 $M_{\odot}$ and eccentricities 0 (solid), 0.2 (dotted), 0.3 (dashed) and 0.4 (dash-dotted). }
\label{lnf-lnt}
\end{figure}

To investigate the nature of the frequency evolution, we make $\ln(f)-\ln(t_c-t)$ plot of the same which is shown in Fig. \ref{lnf-lnt}, where $t_c$ is the time at which the system coalesces. Clearly, binary systems with different eccentricities but same mass components follow similar slope as that of the circular case in the inspiral  regime and thus obeys the power law as $f \propto (t_c-t)^{-p}$ with $p \sim 3/8$.
Subsequently, for a circular binary, the GW frequency in the inspiral phase evolves as,
\begin{equation}
     \ln (f) = C - \frac{3}{8} \ln (t_c-t)
     \label{chirp_mass_ebbh}
 \end{equation}
 where the y-intercept $C \equiv \frac{-3}{8} \ln \left[\frac{256\pi^{8/3}}{5}{\left(\frac{G\mathcal{M}}{c^3}\right)}^{5/3} \right]$ depends on the chirp mass of the system. Fig. \ref{lnf-lnt} shows that binary systems with same chirp mass but different eccentricities have different y- intercepts though similar slopes. This naturally allows us to define a new {\it effective chirp mass $\mathcal{M}_e$} parameter which efficiently captures the inspiral frequency evolution of the non-eccentric (2,2) mode of a non-spinning eccentric system as 
\begin{equation}
    \frac{96 \pi^{8/3}}{5} {\left(\frac{G\mathcal{M}_e}{c^3}\right)}^{5/3} (t_c-t) - \frac{3}{8} f^{-8/3}=0  \,.
    \label{chirpmassE}
 \end{equation}
 This is one of the main results of the paper.
In the following subsection, we develop the phenomenological model for the {\it effective chirp mass} in terms of the chirp mass and the eccentricity of the system.

\section{Phenomenological model for $\mathcal{M}_e$}
\label{model}
In the previous section, we explored the effects of eccentricity in the frequency variation of the signal. In this section we obtain the  parametric model for the new {\it effective chirp mass} parameter $\mathcal{M}_e$ in terms of the numerical fits.

\subsection{Numerical exercise}
\label{NumExer}
 In order to obtain a phenomenological model of $\mathcal{M}_e$ in terms of $\mathcal{M},e$, we consider  stellar black holes having different chirp masses and eccentricities. The lower cut off frequency is chosen to be $10 Hz$. We consider systems such that at least sufficient (one sec and above) duration of the signal falls in the frequency band. As the eccentricity increases the signal duration becomes shorter, hence we restrict up to an eccentricity of 0.6. We consider $\sim$ 300 equal mass systems with $(\mathcal{M},e_{10})$ between $(5-35 M_{\odot}, 0-0.6)$ respectively. The step size was chosen ensuring convergence in the numerical solution.
 
For each system, using the y-intercept of the $\ln f- \ln (t_c-t)$ plot and using Eq. \ref{chirpmassE}, we obtain $\mathcal{M}_e$ as
\begin{equation}
\mathcal{M}_e = \frac{c^3}{G}\left[\frac{5}{256 \pi^{8/3}}  \exp\left(-\frac{8}{3} C\right)\right]^{3/5} .
\end{equation} 
 
\subsection{Polynomial fits for {\it effective chirp mass} $\mathcal{M}_e (\mathcal{M},e_{10}) $}
\label{phenomenological fits}
The GW frequency evolution of eccentric systems varies with $e^2$ according to Eqn \ref{Peters and Mathews equations}. From Sec. \ref{ft-evolution}, it is clear that the {\it effective chirp mass} $\mathcal{M}_e$
depends on $\mathcal{M}$ and the eccentricity $e_{10}$. 
Motivated from this, we use the $\mathcal{M}_e$ value from the numerical exercise and express it in powers of $e_{10}^2$ as,
\begin{eqnarray}
  {\mathcal{M}_{e}} &=& {\mathcal{M}} \bigg( 1 + \alpha(\mathcal{M}) {e_{10}}^2 + \beta(\mathcal{M}) {e_{10}}^4 + \gamma(\mathcal{M}) {e_{10}}^6 \bigg) \nonumber \\
  &\equiv & {\mathcal{M}} {\mathcal G}(e_{10}) .
  \label{phenomenological model}
\end{eqnarray}
The parameters $\alpha(\mathcal{M}),\beta(\mathcal{M}),\gamma(\mathcal{M})$ are numerically computed and fitted to the polynomials of $\mathcal{M}$ as,
\begin{subequations}

\begin{align}
\alpha(\mathcal{M}) &= \xi {\bigg(\frac{\mathcal{M}}{M_{\odot}}\bigg)} + \delta  , \label{alpha}&\\ 
    \beta(\mathcal{M}) &= \Xi_{\beta} {\bigg(\frac{\mathcal{M}}{{M_{\odot}}}\bigg)}^2 + \Delta_{\beta} {\bigg({\frac{\mathcal{M}}{{M_{\odot}}}}\bigg)}^4 + \kappa_{\beta} {\bigg({\frac{\mathcal{M}}{{M_{\odot}}}}\bigg)}^6  \nonumber \\
    &+  \zeta_{\beta}  {\bigg(\frac{\mathcal{M}}{{{M_{\odot}}}}\bigg)}^8  , \label{beta}  \\
    \gamma(\mathcal{M}) &= \Xi_{\gamma} {\bigg(\frac{\mathcal{M}}{{M_{\odot}}}\bigg)}^2 + \Delta_{\gamma} {\bigg(\frac{\mathcal{M}}{{{M_{\odot}}}}\bigg)}^4 + \kappa_{\gamma} {\bigg(\frac{\mathcal{M}}{{{M_{\odot}}}}\bigg)}^6  \nonumber \\
    &+  \zeta_{\gamma} {\bigg(\frac{\mathcal{M}}{{{M_{\odot}}}}\bigg)}^8 \label{gamma} 
\end{align}
\end{subequations}

\begin{table} [htb]
\begin{tabular}{c c}
\hline\\
Parameter & Value \\
\\
\hline\\
$\xi$  & 0.06110974175360381 \\
$\delta$ & -0.4193723077257345\\
$\Xi_{\beta}$  & 0.00801015132110059  \\
$\Delta_{\beta}$  & $-2.14807199936756\times10^{-5}$ \\
$\kappa_{\beta}$  & $1.12702400406416\times10^{-8}$ \\
$\zeta_{\beta}$ & $-1.9753003183066\times10^{-12}$\\
$\Xi_{\gamma}$ & 0.024204222771565382  \\
$\Delta_{\gamma}$& $-6.261945897154536\times10^{-6}$ \\
$\kappa_{\gamma}$ & $1.1175104924576945\times10^{-8}$ \\
$\zeta_{\gamma}$ & $-3.681726165703978\times10^{-12}$\\
\hline
\end{tabular}
\caption{Fit parameters of the phenomenological model}
\label{parameter values table}
\end{table}
The values of the fitted parameters are given in the Table \ref{parameter values table}.
The $\chi^2$ error for the fits varies between $O(10^{-2})$ to $O(10^{-3})$. Including terms beyond powers of $e_{10}^6$ increases the $\chi^2$ error to $O(10^{-1})$. Hence we restrict upto $e_{10}^6$ terms.
Clearly, for circular binaries $e_{10}=0$, Eq. \ref{phenomenological model} reduces to $\mathcal{M}_e=\mathcal{M}$.

\begin{figure}
    \includegraphics[scale=0.5]{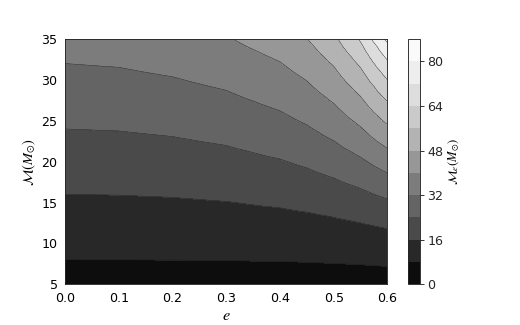}
    \caption{{\it Effective chirp mass} variation in the chirp mass and eccentricity plane. The shaded regions represent iso-$\mathcal{M}_e$ contours. The eccentricity is the eccentricity at 10 Hz viz. $e_{10}$.}
    \label{mc e mce plot}
     \end{figure}
Fig \ref{mc e mce plot} shows the variation of $\mathcal{M}_e$ in the chirp mass and $e_{10}$ plane. The colour bar shows the $\mathcal{M}_e$ of the system obtained from the phenomenological model. The shaded regions represent iso-$\mathcal{M}_e$ contours. For low chirp mass, the $\mathcal{M}_e$ does not change significantly with eccentricity and hence the circular chirp mass is sufficient to explain the dynamics of the system. However, with the increase of chirp mass, the eccentricity driven dynamics is governed by $\mathcal{M}_e$ which becomes more and more prominent for higher eccentricities.  

\begin{figure}
\hspace*{-1.cm}
    \includegraphics[scale=0.51]{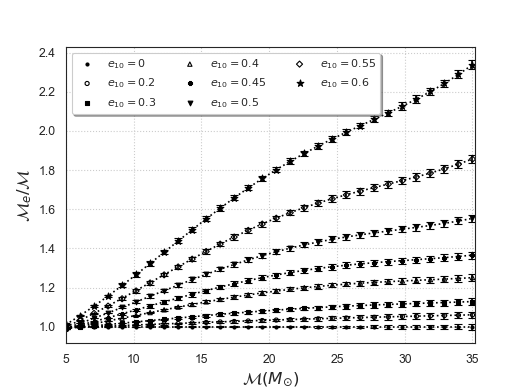}
    \caption{$\mathcal{M}_e/\mathcal{M}$ vs $\mathcal{M}$ for different eccentricities. The points denote a system with different $\mathcal{M},e_{10}$ obtained numerically. The dotted lines are the fitted curves. }
    \label{McebyM diff e}
     \end{figure}

Fig \ref{McebyM diff e} shows the $\mathcal{M}_e/\mathcal{M}$ vs $\mathcal{M}$ variation for binary systems with different values of $e_{10}$. The curves with different markers correspond to different eccentricities.
For low eccentricities up to $e_{10}<$0.3, the $\mathcal{M}_e$ agrees up to 90$\%$ with $\mathcal{M}$. As the eccentricity increases, the {\it effective chirp mass} deviates from the chirp mass. This is because higher the eccentricity, the system tries to radiate energy at the higher rate and thus giving higher $\mathcal{M}_e$. For eccentricity as high as 0.6 and beyond $\mathcal{M} = 25 M_{\odot}$, the value of $\mathcal{M}_e$ becomes more than twice of $\mathcal{M}$. This effect will be prominent in third generation interferometers where the stellar mass compact binaries are expected to have longer inspiral signals owing to the lower frequency cut-off values. Here, we develop the model in terms of $\mathcal{M}$ and $e_{10}$ eccentricity with a reference frequency of 10 Hz. If the same binary system is observed in advanced detectors such as Cosmic Explorer \cite{reitze2019cosmic} , Einstein Telescope \cite{Maggiore_2020} etc., the $f_L$ will be 1 Hz but value of $\mathcal{M}_e$ should be the same. Thus, if one wants the model in terms of $f_L=1$Hz then the functional form of the $\mathcal{M}_e$ can be different {\it i.e.} $\mathcal{M}_e = \mathcal{M} {\mathcal{G}}({e_{10}}) = \mathcal{M} p(e_1)$. Alternatively, one can still use the same model with the reference frequency of 10 Hz under the given assumptions. In this paper we will use the reference eccentricity to be defined at 10 Hz viz. $e_{10}$, and henceforth for convenience we use $e$ and $e_{10}$ interchangeably.
\section{Eccentric non-spinning inspiraling binary in the advanced detector}
\label{Eccentric non-spinning inspiraling binary in the advanced detector}
In this section, we study the loudness of the inspiral phase of the eBBH system. Here, we consider the upgraded configuration of the advanced interferometric detectors with Aplus noise power spectral density for better detection prospects of the inspiral phase of an eccentric binary system [36].  We consider the Aplus noise curve with a lower cut-off frequency of 10 Hz.

We consider non-spinning, eBBH systems with source frame component masses $m_{1,2}$ between 5 $M_{\odot}$ and 100 $M_{\odot}$ for this injection study. We sample the primary component mass from a power law distribution $p(m_1) = {m_1}^{-\alpha}$ with index $\alpha =2.3$ and secondary mass as uniform between $(5M_{\odot}, m_1)$ \cite{GWTC1} . We distribute the binary sources uniformly in comoving volume with luminosity distance between 100 Mpc to 1 Gpc. We place the sources uniformly over the celestial sphere, the inclination angle and other parameters are distributed uniformly. We simulate eccentric signals using EccentricTD waveform, sourced using LALSuite software \cite{lalsuite}, for the simulations. The waveform models the inspiral phase of the eccentric binary system using TaylorT4 approximant up to 2PN order. Following the waveform constraints, we sample the eccentricity from a uniform distribution between 0.01 and 0.6. We compute the optimal SNR of the system using the Aplus noise curve. 

\begin{figure}[thb!]
    \includegraphics[scale=0.5]{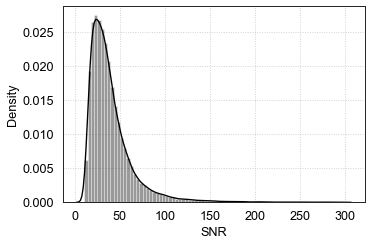}
    \caption{Optimal SNR distribution of eBBH systems using EccentricTD waveform and Aplus noise curve.}
    \label{SNR distribution}
     \end{figure}

Fig \ref{SNR distribution} shows the SNR distribution of the simulated signals. The distribution has a mean SNR value around 40. Around 44$\%$ of the injections have SNR value greater than 40. Hence, given an astrophysically motivated scenario, we expect a good number of systems with sufficient inspiral SNR in the advanced detector band and a good possibility that few of these loud systems are eBBH systems.

\begin{figure}[thb!]
\hspace*{-0.8cm}
    \includegraphics[scale=0.5]{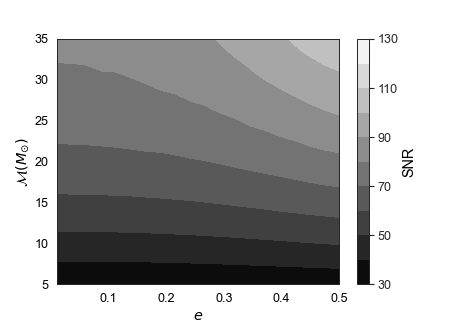}
    \caption{Optimal SNR in the chirp mass and eccentricity plane. The shaded regions represent iso-SNR contours.}
    \label{mc e snr plot}
     \end{figure}

In Figure \ref{mc e snr plot}, we study the variation of the inspiral SNR in the chirp mass and eccentricity space. Here, we fix all the parameters except chirp mass and the eccentricity. 
We place the systems at the distance of 400 Mpc. The colour bar shows the optimal SNR of the system. We note that higher is the chirp mass as well as eccentricity, shorter is the signal duration but more is the signal power emitted, see Eq. \ref{dE_dt}.  Thus, the system with higher chirp mass and eccentricity gives higher value of inspiral SNR. The SNR trend clearly follows the {\it effective chirp mass} trend as shown in Fig. \ref{mc e mce plot} at least in the region of low chirp masses. 

\section{Implication on the chirp mass obtained from $f-t$ morphology}
\label{MeFromQtransform}
In the previous section, we obtain the phenomenological model for the {\it effective chirp mass} parameter for the non-spinning eccentric system. In this section, we estimate this parameter and state the implication of this in the unmodelled searches. 

\subsection{$\mathcal{M}_e$ and Q-transform representation}

In section \ref{Gravitational wave frequency evolution of non-spinning eccentric system}, we show that the {\it effective chirp mass parameter} of the system captures the frequency-time morphology of the lowest (2,2) mode of the gravitational wave signal from the non-spinning eccentric binary system. Projecting the GW strain signal in the time-frequency representation provides a natural way to estimate this parameter. There are numerous quadratic transforms for the same such as spectrogram, scaleogram, Q-transform, etc. Here, we use the Q-transform \cite{Chatterji_2004}. The Q-transform, $X(t,f_0,Q)$, is a windowed Fourier transform that projects the over-whitened time-series data onto a multi-resolution time-frequency basis. $f_0$ is the central frequency and $Q$ is the Q-tile in the Q-transform. The energy of a Q transformed signal at a time-frequency-Q location(pixel)  is proportional to $\|X(t,f_0,Q\|^2$. For further details, please refer to Appendix \ref{Q transform}.

\begin{figure}[t!]
\hspace*{-1cm}
\includegraphics[scale=0.5]{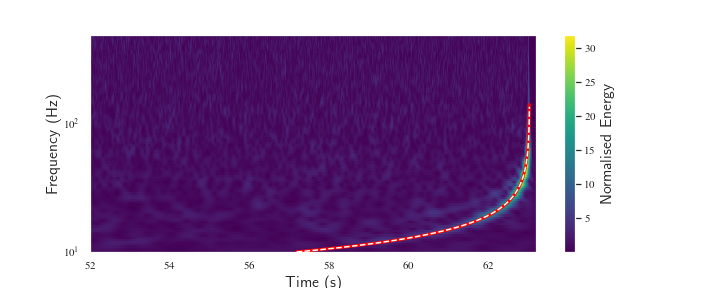}\\
\caption{Q transform of a 30-30 $M_{\odot}$ system with $e_{10}=0.1$ using EccentricTD waveform. The dashed white line correspond to the $f(t)$ curve obtained by evolving the numerical equations. The solid red curve corresponds to 
the $f(t)$ evolution from the phenomenological model of ${\mathcal M}_e$ from Eqn \ref{chirpmassE}.}
\label{Q transform of ebbh system introduction}
\end{figure}

Fig \ref{Q transform of ebbh system introduction} shows the Q transform of a 30-30 $M_{\odot}$ system with $e_{10}=0.1$ generated using EccentricTD waveform. The signal is represented by a chirp track in the time-frequency representation with bright pixels carrying the frequency evolution information. The dashed line corresponds to the $f(t)$ curve obtained by evolving the numerical equations. The solid red curve corresponds to 
the Eq. \ref{chirpmassE} using the {\it effective chirp mass} 
${\mathcal M}_e$ model given by Eq. \ref{phenomenological model}. All three curves show good overlap between them. 
The second fainter track corresponds to the higher eccentric harmonic. For low to moderate eccentricities, these harmonics do not have a significant contribution, but for moderate to high eccentricities, the power in these harmonics has a non-negligible contribution with respect to the lowest (2,2) track which we model here.

\subsection{$\mathcal{M}_e$ estimation}

The Q-transform provides a time-frequency representation of the lowest chirp track $f(t)$. Here, we use the same to estimate the {\it effective chirp mass} parameter of the underlying signal. For the same, we follow the intuitive approach laid down below.

We run the simulation with GW signal from eccentric systems injected in the Gaussian noise coloured by Aplus noise power spectral density. We take the following steps: We obtain the Q-transform of the simulated data. The best fit Q value is chosen from a range of Q values based on maximum energy. 
We extract the bright energy pixels along the chirp track by putting an energy cut on the pixel energy. The signal energy is proportional to the SNR, and its distribution in to different pixels depends on the injected $\mathcal{M}$ and $e_{10}$ of the system. We apply a simple procedure to extract pixels from the lowest quadrupolar (2,2) track. We consider pixels with energy threshold cuts between $6 \sigma$ to $14 \sigma$ values where $\sigma$ is determined from the pixel energy distribution. For each threshold value, we use the extracted pixels to estimate the {\it effective chirp mass} $\mathcal{M}_e$ of the system as described in section \ref{NumExer}. We choose that energy cut as the final which corresponds to that $\sigma$ threshold which gives the maximum value of estimated $\mathcal{M}_e$ amongst all the values. We denote this as ${\mathcal{M}_e}^{est}$. 

The range of $\sigma$ values are chosen to address the range of masses. For low chirp mass and low eccentricity cases, the duration of the signal is higher in the detector band. For high chirp mass and high eccentricity cases, the duration of the signal is shorter and along with that the eccentricity harmonics appear. For higher eccentricities (above 0.4), energy contribution in the higher eccentric harmonics is higher \cite{peters}. We take systems with eccentricity below 0.4 to minimise that effect. Low energy cut gives stray pixels while high energy cut gives fewer pixels from the desired chirp track. The above criterion gives an optimum choice of energy cut.

\begin{figure}[h!]
\hspace*{-1cm}
    \centering
    \includegraphics[scale=0.5]{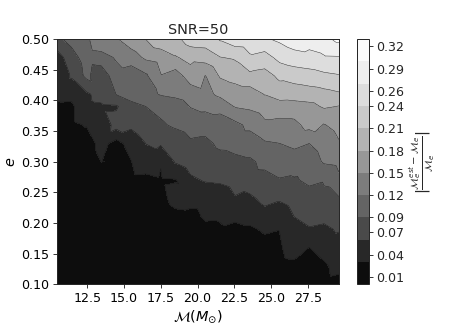}
    \caption{Estimation of $\mathcal{M}_e$: relative error between the estimated $\mathcal{M}_e$ and that from the phenomenological model in the chirp mass eccentricity plane for the fixed optimal SNR of 50.}
    \label{relative error}
     \end{figure}

Fig \ref{relative error} summarises this study for the non-spinning eccentric systems with fixed SNR of 50.
The image grid corresponds to injected systems in the $\mathcal{M},e$ space and the colour bar represents the relative error between the estimated ${\mathcal{M}_e}^{est}$ and that of the phenomenological model ${\mathcal{M}_e}$. We observe that for low chirp masses, the $\mathcal{M}_e$ can be estimated accurately up to moderate eccentricities. As the $\mathcal{M}$ as well as eccentricity increases, the duration of the signal decreases and hence decreases the accuracy in estimation of ${\mathcal{M}_e}$ due to small number of pixels recovered. Allowing as high as $10 \%$ error  in the estimation of ${\mathcal{M}_e}$, we can reliably estimate ${\mathcal{M}_e}$ up to 0.5 initial eccentricity for low chirp mass and up to initial $e\sim0.25$ for $\mathcal{M}\sim30 M_{\odot}$. 

Current burst searches use chirp mass $\mathcal{M}$ cut to constrain the signal morphology and reconstruct the signal \cite{drago2021coherent}. For eccentric systems, this cut needs to be tuned and as shown above, the pixels should be selected based on the ${\mathcal{M}_e}$ cut and not the ${\mathcal{M}}$ cut. Incorporating this information in the eBBH burst searches is crucial in future searches even for moderate eccentricities.

\section{Implication on constraining the eccentricity}
\label{Implication on the chirp mass and eccentricity estimation}
After the confident detection of an astrophysical GW signal, typically  we use Bayesian inference framework \cite{Thrane_2019} to estimate the parameters of the underlying compact binary system. However, due to lack of reliable models, as well as systematic bias in parameter estimation, we investigate if the {\it effective chirp mass} model proposed here can provide a handle to constrain the initial eccentricity in the model independent way.

For a given system, once we have estimated  
$\mathcal{M}_e$ from the time-frequency representation, we obtain the rough location of the chirp mass from Figure \ref{mc e mce plot} by placing $10\%$ uncertainty on the chirp mass and exploring range of eccentricity (0,0.5). We set up a (0.08,0.008) grid in the $(\mathcal{M},e)$ space. For each point in this grid, we obtain the $f(t;{\mathcal M}_e)$ curve using the {\it effective chirp mass} model given by Eq. \ref{chirpmassE} and \ref{phenomenological model}. We collect the pixels along the neighbourhood of the chirp track $f(t;{\mathcal M}_e)$ and evaluate the total energy of the collected pixels
\begin{equation}
   E \sim \sum_{k} \|X_k\|^2,
\end{equation}
where $k$ denotes the $k$-th pixel.
We represent the energy in the $(\mathcal{M},e)$ plane in the form of a energy heat map.
For high inspiral SNR values, that $(\mathcal{M},e)$ combination which gives the maximum energy amongst all the grid-points should give the required $(\mathcal{M},e)$ pair. However, the $f(t;{\mathcal M}_e)$ is a function of $\mathcal{M}_e$ and hence harbours an inherent degeneracy in the $(\mathcal{M},e)$ space. 

We break this degeneracy by employing the Bayesian parameter estimation approach to estimate the chirp mass $\mathcal{M}$ using the quadrupolar non-spinning quasi-circular waveforms. We use this error estimated from the parameter estimation study on the chirp mass to constrain the chirp mass. Finally, we combine this information with the energy heat-map obtained using the {\it effective chirp mass} model to constrain the eccentricity of the system.

For validation, we choose an equal mass binary black hole system with chirp mass $28 M_{\odot}$ and initial eccentricity $0.15$ respectively. We simulate the GW signal from this system using EccentricTD waveform and inject into simulated Gaussian noise with Aplus power spectral density. The optimal SNR of the system was chosen to be around 50. The effective chirp mass estimated was $27.61 M_{\odot}$. We choose the chirp mass range to be $24 M_{\odot}-32M_{\odot}$ and eccentricity range between $0.01-0.42$.

\begin{figure}
    \includegraphics[scale=0.52]{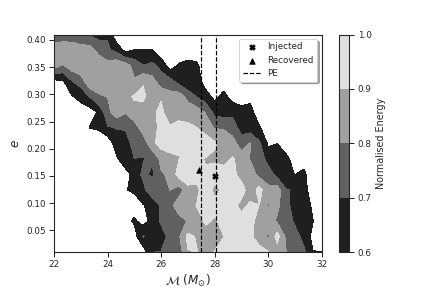}
    \caption{Heatmap of injected system with $(\mathcal{M},e)=28 M_{\odot},0.15$. The regions with different colours represent iso-energy contours. }
    \label{heatmap}
     \end{figure}

Using the $\mathcal{M}-e$ grid, we compute the time-frequency chirp tracks using Eq. \ref{chirpmassE}. We select the pixels in the neighbourhood of those tracks and evaluate the total energy. In Fig \ref{heatmap}., we show the energy heat map with a maximum value set to unity in the 
$(\mathcal{M},e)$ plane. The different levels in the energy contours represent drop in the energy from the maximum as the $(\mathcal{M},e)$ grid point moves away from the maxima. We note that the maxima(triangle point) corresponding to the total energy is not sharp but the drop in energy is gradual. This is expected as we are not explicitely considering the signal phasing information but only the time-frequency morphology in terms of the chirp track.

For the same injection, we perform parameter estimation using BILBY package \cite{Romero-Shaw:2020owr} using IMRPhenomPv2 waveform by setting spin to zero value. We put a uniform prior on chirp mass ranging from $5M_{\odot}-60M_{\odot}$. 
We use the DYNESTY sampler \cite{Speagle_2020} which is based on the nested sampling algorithm implemented in the BILBY. The black dotted lines show the $3 \sigma$ credible interval around the median value of chirp mass of $27.78 M_{\odot}$ recovered from the parameter estimation. Using the results from the parameter estimation study and the figure \ref{heatmap}, we place the constrain on the eccentricity of the system to be $0.23$ at the $90\%$ of the maximum total recovered energy value. 

\section{Conclusions}
\label{conclusions}
We know that the frequency evolution of the
GW for the quadrupolar (2,2) mode of the quasi-circular compact binary in the inspiral phase is governed by the chirp mass parameter. For eccentric systems, the rate of energy emission depends on the asymmetry in the orbit governed by the eccentricity parameter and thus affects the frequency evolution. In this work we show that one can capture the frequency evolution of the non-eccentric part of the (2,2) mode for the non-spinning eccentric system with an equivalent {\it effective chirp mass} parameter; $\mathcal{M}_e$ which depends on the chirp mass and the initial eccentricity of the system.

We develop a phenomenological model for $\mathcal{M}_e$ in terms of the chirp mass and the initial eccentricity at a reference frequency of 10 Hz. The proposed model is developed for chirp mass up to $30 M_{\odot}$ and the initial eccentricity approximately up to 0.6 for the advanced detectors. We clearly demonstrate that the GW frequency evolution of the (2,2) mode (excluding the eccentric harmonics) during the inspiral phase is similar to that of its quasi-circular counterpart by just replacing the chirp mass with the newly defined {\it effective chirp mass} for non-spinning eccentric systems. We note that while matched filter based templates require more detailed phase evolution of the eccentric signal which include post-Newtonian corrections, the model independent excess power searches can gain from this phenomenological model to constrain the eccentricity of the non-spinning eBBH systems in the advanced GW detectors.

We provide two direct implications of our proposal in GW astronomy. {\it First}; in model independent GW searches chirp-cut is one important attribute to distinguish between compact binaries of different mass ranges \cite{Abbott_2019}. This model has direct impact on implication on the eccentricity after estimating the chirp mass from the pixels. {\it Second}; once we estimate the {\it effective chirp mass parameter} using the time-frequency representation, the model can be used along with its proposed analytical frequency evolution and the chirp mass parameter estimation using the quasi-circular, non-spinning waveforms to constrain the eccentricity of the system especially for the loud eBBH systems up to moderate values of eccentricities.


We provide a proof of principle approach of the same by simulating few examples of equal mass, non-spinning eBBH systems using EccentricTD waveform in the simulated Gaussian noise of the Advanced LIGO detector with the Aplus noise power spectral density. We have restricted ourselves to low and moderate values of eccentricity in the effective chirp mass estimation. This is attributed to the fact that with an increase of eccentricity, the higher eccentric harmonics have a non-negligible contribution compared to the lowest (2,2) track, which results in picking up pixels from those tracks along with the (2,2) track. Hence the effective chirp mass estimation suffers, which in turn leads to increased error in the eccentricity constraints. In addition to this, for asymmetric mass systems with non-circular orbits, higher-order modes also appear due to mass asymmetry. Thus, the problem will be further complexified with higher harmonics contributing to both mass asymmetry as well as eccentricity. We shall address these problems in a future work.

In summary, the GW astronomy has grown rapidly in last five years. While we have detected several quasi-circular binary systems, no confirmed detection of eccentric system so far. Limitations due to available waveforms constraint the conventional model based methods in searched as well as estimation in the signal parameters. Astrophysical models predict finite chance to observe eBBH systems in the dense environment. In near future generation of detector sensitivity where we will have more number of detection out of which a significant number will have non negligible eccentricity. Here, we propose a new,
{\it effective chirp mass} parameter which captures the frequency evolution of the (2,2) non-eccentric mode of the non-spinning eccentric binary evolution. We indicate some direct implications of this work in the model independent approaches such as constraining the eccentricity of a loud, non-circular, non-spinning binary system.

\section{Acknowledgements}
The authors acknowledge Srishti Tiwari, K.G.Arun, Achamveedu Gopakumar, Maria Haney,  Chandra Kant Mishra, Ravikiran Hegde and Ganesh Rohan for providing useful comments and suggestions for the manuscript.  The authors thank the anonymous referee for useful comments. NB acknowledges Inspire division, DST, Government of India for the fellowship support. AP acknowledges the SERB Matrics grant MTR/2019/001096 and SERB-Power-fellowship grant SPF/2021/000036 of Department of Science and Technology,  India for support.  The authors thank the LDG clusters for the computing resource. This document has LIGO DCC No P2100280. 
\textit{We want to thank all of the essential workers who put their health at risk during this ongoing COVID-19 pandemic. Without their support, we would not have completed this work. We offer condolences to people who have lost their family members during this pandemic.}

\bibliographystyle{apsrev4-1}
\bibliography{ref}

\appendix

\section{Instantaneous semi-major axis in terms of eccentricity $a(e)$}
\label{appendix B}
Using Eqn \ref{da_dt} and \ref{de_dt}, we get,
\begin{equation}
    \frac{da}{de}=\frac{12~a}{19} \frac{[1+\frac{73e^2}{24}+\frac{37e^4}{96}]}{e(1-e^2)[1+\frac{121e^2}{304}]}
    \label{da_de} \,.
    \end{equation}
Eqn \ref{da_de} can be integrated to give \cite{Maggiore_2020},
\begin{equation}
    a(e) = C_0 \frac{e^{12/19}}{(1-e^2)}\bigg[1+{\frac{121 e^2}{304}}\bigg]^{870/2299} \equiv C_0 g(e) \,,
    \label{aofe}
\end{equation}
where $C_0$ can be determined from the initial condition that, $a=a_{0}$ when $e=e_{0}$ where $e_{0}$ is the eccentricity when the GW signal enters the detector band with GW frequency coinciding with the lower cut-off frequency $f=f_L=10 Hz$. 
Hence $a_{0}=C_0 g(e_{0})$. Now using Kepler's third law and considering the fact that we look at non eccentric (2,2) harmonics, we use the GW frequency to be twice the orbital frequency.  Hence we have   $\pi^2 a_{0}^3~f^2=GM $, we obtain $a_{0}$ as,
\begin{equation}
    a_{0}=(GM)^{1/3}\pi^{-2/3}f_L^{-2/3} \,,
    \label{a_0}
\end{equation}
where $M$ is the total mass of the system.
Hence using Eqn \ref{aofe} , \ref{a_0} ,and $C_0$, we write,
\begin{equation}
    a(e)=\frac{(GM)^{1/3}\pi^{-2/3}f_L^{-2/3}}{g(e_{0})}\frac{e^{12/19}}{(1-e^2)}\bigg[1+{\frac{121 e^2}{304}}\bigg]^{870/2299} \,.
\end{equation}

\section{Q transform}
\label{Q transform}
The Q-transform is defined as a windowed Fourier transform that  projects the over-whitened time-series (or frequency series) onto a multi-resolution time-frequency basis . In a way, it is quite similar to a continuous wavelet transform. Q-transform, $X(t,f_0,Q)$ of an  over-whitened frequency series with an integral operator with kernel  $\mathfrak{\tilde{w}^*}(f,f_0,Q)e^{i2\pi f t}$, as (\textcolor{green}{\cite{Chatterji_2004}}):
\begin{equation}
\begin{split}
    X(t,f_0,Q) 
    &= \int_{-\infty}^{\infty} df ~ \tilde{\tilde{x}}^*(f+f_0)\mathfrak{\tilde{w}}(f,f_0,Q)e^{i2\pi f t}
    \end{split}
    \label{Qtransform eqn}
\end{equation}
Here the over-whitened data given as $\tilde{\tilde{x}}(f) = \tilde{x}(f)/S_n(f)$ with $\tilde{x}(f)$ as the frequency domain representation of $x(t)$ and $S_n(f)$ is the two sided noise power spectral density (PSD) of the interferometer noise. $f_0$ is the central frequency and $Q$ is the Q-tile used to calculate the Q-transform.

The window function in frequency domain $\mathfrak{\tilde{w}}(f,f_0,Q)$ is a normalised bi-square window with finite frequency domain support centered at zero as \textcolor{green}{\cite{shouravthesis}} :
\begin{equation}\label{eq:2}
    \mathfrak{\tilde{w}}(f,f_0,Q) = \begin{cases}
\sqrt{\frac{315Q}{128\sqrt{11}f_0}}\Big[1 - \Big(\frac{fQ}{f_0\sqrt{11}}\Big)^2 \Big]^2, \hfill \|f\| \leq \frac{f_0\sqrt{11}}{Q} \\
\\
0, \hfill otherwise
\end{cases}
\end{equation}

\end{document}